\documentclass[aps,prd,preprint,superscriptaddress,nofootinbib]{revtex4-2}

\usepackage{amsmath}
\usepackage{amssymb}
\usepackage{mathtools}
\usepackage{tensor}
\usepackage[dvipsnames]{xcolor}
\usepackage{hyperref}
\usepackage{bm}
\usepackage{graphicx}
\usepackage{caption}
\usepackage{subcaption}
\usepackage{setspace}
\usepackage{lipsum}

\newcommand{\order}[1]{\mathcal{O}\!\left(#1\right)}
\renewcommand{\Re}[1]{\text{Re}\!\left(#1\right)}
\renewcommand{\Im}[1]{\text{Im}\!\left(#1\right)}
\newcommand{\abs}[1]{\left| #1 \right|}
\newcommand{\magsq}[1]{\left| #1 \right|^2}
\newcommand{\vk}{\vec{k}}
\newcommand{\masslesslimit}{\lim\limits_{m\to 0}}
\newcommand{\gI}{g^{I}_k}
\newcommand{\gII}{g^{II}_k}

\newcommand{\gIIc}{g^{II*}_k}
\definecolor{linkblue}{RGB}{11,95,217}
\hypersetup{colorlinks,
	allcolors = linkblue}

\begin{document}
\title{Argument for the radiation-dominated behavior of matter fields in the preinflationary era}
\author{Taylor M. Ordines}
\author{Eric D. Carlson}
\affiliation{Department of Physics, Wake Forest University, 1834 Wake Forest Road, Winston-Salem, North Carolina 27109, USA}
\date{\today}

\begin{abstract}
  We investigate the leading-order behavior of matter fields in the preinflationary era using the semiclassical approximation.
  Many inflationary models assume without supporting arguments that the Universe was radiation dominated prior to inflation, leading to modifications of cosmological observables, such as the Cosmic Microwave Background power spectrum.
  In previous work, we demonstrated that conformally coupled scalar fields do have a radiation-like contribution to the stress-energy tensor at sufficiently early times.
  In this work, we extend these arguments to apply to massless spin-1 fields and massive or massless spin-$\frac{1}{2}$ fields.
  We find massless spin-1 fields always have a radiation-like contribution.
  For spin-$\frac{1}{2}$ fields, we find the contribution at early times is radiation-like assuming this is the dominant contribution to the stress-energy tensor.
\end{abstract}

\maketitle
\newpage
%\tableofcontents

\section{Introduction}
The inflationary paradigm explains many characteristics of our observed Universe.
Many models of inflation make use of a radiation-dominated behavior of matter fields in the preinflationary era in order to obtain modifications to the standard predictions of inflation and better explain observed phenomena (see~\cite{das_revisiting_2015} and references therein for a review of models using a radiation-dominated preinflationary era).
For instance, various models involving the transition from a radiation-dominated era to inflation lead to modifications of the Cosmic Microwave Background anisotropy spectrum, such as a lowering of the quadrupole moment which appears to be anomalously suppressed~\cite{bennet_first_2003}.
The assumption that fields are radiation-dominated prior to inflation is central to such models, though it is non-trivial to demonstrate that such behavior was the case for our Universe, and it would be hopeless to try to detect particles due to these fields today given the effects of inflation.
It is therefore interesting to analyze the preinflationary era and the behavior of matter fields in it.

The primary objective of this work is to investigate the behavior of quantum fields in the preinflationary era.
A complete analysis of the the early universe would require a theory of quantum gravity, which for now is out of reach, but one could anticipate that after the Planck era some span of the preinflationary era would have curvature well below the Planck scale, in which case the semiclassical approximation should hold.
In this paper, we will assume the semiclassical approximation to hold in some portion of the preinflationary era, and we will use this framework to analyze quantum fields in a Friedmann-Lema\^{i}tre-Robertson-Walker (FLRW) spacetime and obtain their corresponding contributions to the energy density.

The case of massive scalar fields was investigated using the semiclassical approximation in~\cite{anderson_semiclassical_2020}.
There it was shown that under a set of conditions on the scale factor the fields did result in radiation-dominated contributions to the energy density.
The conditions on the scale factor were also argued to likely be valid for our Universe.
In the present work, we will investigate the corresponding behavior of other fields, namely massless spin-1 and massive or massless spin-$\frac{1}{2}$ fields.

The analysis for spin-1 and spin-$\frac{1}{2}$ fields is generally more complicated than for scalar fields.
For spin-$\frac{1}{2}$ fields, the mass appears explicity in the expression for the counterterms at higher than zeroth adiabatic order, which leads to additional complications.
Performing adiabatic regularization with a zeroth-order parameterization of the states does not explicitly appear to produce finite energy densities until one considers the higher order contributions buried in the parameterization. For an overview of the adiabatic regularization procedure for spin-$\frac{1}{2}$, see~\cite{landete_adiabatic_2013,landete_adiabatic_2014,rio_renormalized_2014}.
For spin-1 fields one must consider the massive and massless cases separately.
Massive vector fields do appear in the Standard Model, but they are ultimately due to interactions with the Higgs field and are out of the scope of this paper.
For the massless case one can show that the analysis decomposes into four decoupled copies of a scalar field, as was done in Ref.~\cite{chu_adiabatic_2017}.
There the analysis was performed to obtain the trace anomaly, but obtaining the renormalized energy density from this groundwork is non-trivial.

In the following, we work in Planck units with $c=G=\hbar=1$ and use the $(-,+,+,+)$ signature for the metric.
We assume a spatially flat universe described by the FLRW metric
\begin{equation}
  ds^2 = a^2 \left( -d\eta^2 + d\vec{x}^2 \right) \;,\label{eq:FLRWMetric}
\end{equation}
where $a\equiv a(\eta)$ is the scale factor and $\eta$ is conformal time.
We will employ a prime, such as $a'\equiv \partial_\eta a$, to denote differentiation with respect to conformal time.
We are interested in taking $a\rightarrow 0$ at early times and determining whether one can assume radiation-dominated behavior.
However, we do not want to consider times in the Planck era, during which the semiclassical approximation is not assumed to be valied, so we denote by $\eta_0$ the earliest conformal time of consideration and implicitly assume it to be past the Planck era.
In Planck units, this corresponds to a value of the Hubble parameter $H\equiv\frac{a'}{a^2}\ll 1$.
We therefore aim to determine if the renormalized energy density $\rho_r$ for each field is appropriately radiation dominated during a range of conformal time $\eta_0<\eta<\eta_1$.
We will allow for the possibility of a non-zero mass $m$ for fermions, in which case we will insist that $m\ll 1$ in Planck units.
If there are fermion fields with $m\gtrsim 1$, these will have mass comparable to or greater than the Planck mass, and then we will assume the fields have no contribution to the total energy density when $\eta\gtrsim\eta_0$.

The body of this article is split into three sections, one for each of the two spins of fields and one for a discussion of the results obtained.
Section~\ref{sec:VectorFields} contains the analysis for spin-1 vector fields, beginning with a summary of the adiabatic regularization procedure and concluding with the renormalization of the energy density and our main result for the spin-1 case.
Section~\ref{sec:SpinorFields} contains the analysis for spin-$\frac{1}{2}$ fields and begins with a summary of the modified adiabatic regularization procedure, as described in~\cite{rio_renormalized_2014}.
The subsequent renormalization and analysis is more complicated than that for the vector fields and is split into subsections: in Sec.~\ref{sec:SpinorFieldsRenormalizedEnergyDensity} we give the renormalization counterterms and preview the assumptions built into our analysis; in Sec.~\ref{sec:SpinorFieldsBounds}, we derive bounds on the renormalized energy density by splitting the contributions into high and low energy terms and analyzing each in turn; and in Sec.~\ref{sec:SpinorFieldsFriedmann} we obtain the leading order behavior of the renormalized energy density and our main result for the spin-$\frac{1}{2}$ case.
We close in Sec.~\ref{sec:DiscussionAndConclusion} with a discussion of our main results and final remarks.

\section{Energy Density for Vector Fields\label{sec:VectorFields}}
As discussed in the introduction, massive and massless vector fields must be treated separately.
In the Lagrangian description, the massive field is described by the Proca action~\cite{schambach_proca_2018}, which in flat spacetime is a gauge-fixed theory involving the Higgs mechanism.
Working with interactions and the curved spacetime form of the action is beyond the scope of this paper.
We will focus instead on the massless case.

According to an argument in~\cite{wald_1978}, any conformally invariant theory in a flat FLRW spacetime will have a stress-energy tensor that contains two terms, one of which is radiation-like and the other of which is the anomalous term.
Renormalizing the electromagnatic field, however, requires that masses are introduced for the photon and ghost fields, which break the conformal invariance.
These masses are then taken to zero to obtain physical results.
We therefore feel it is worth working through the details of this procedure to confirm that the argument in~\cite{wald_1978} works.

The massless vector field in curved spacetime is given by the massless limit of the theory described by Lagrangian~\cite{chu_adiabatic_2017}
\begin{equation}\label{eq:VectorFieldLagrangian}
  \mathcal{L} = \sqrt{-g}\left[ -\frac{1}{4}g^{\mu\rho}g^{\nu\sigma}F_{\mu\nu}F_{\rho\sigma} - \frac{1}{2\xi}\nabla^{\mu}A_{\mu}\nabla^{\nu}A_{\nu} - \frac{1}{2}m^2g^{\mu\nu}A_{\mu}A_{\nu} + ig^{\mu\nu}\partial_{\mu}\bar{\chi}\partial_{\nu}\chi + im_\chi\bar{\chi}\chi \right] \;,
\end{equation}
where $A_{\mu}$ is the four-vector, $\xi$ is the gauge fixing parameter, $\chi$ is the (complex) ghost field used to maintain gauge invariance, $m_\chi$ is the mass of the ghost field, and $F_{\mu\nu} = \partial_{\mu}A_{\nu} - \partial_{\nu}A_{\mu}$.
Masses are included in the Lagrangian~\eqref{eq:VectorFieldLagrangian} in order to properly renormalize the theory.
These masses can trivially be taken to zero before computing any unrenormalized observables, but they are crucial in obtaining the appropriate renormalization counterterms.
We will make clear when they are to be taken to zero in the following.
The energy-momentum tensor from Eq.~\eqref{eq:VectorFieldLagrangian} is
\begin{subequations}
  \begin{align}
    T_{\mu\nu} = & T_{\mu\nu}^{\text{Maxwell}} + T_{\mu\nu}^{\text{G}} + T_{\mu\nu}^{\text{ghost}} \;,                                                                                                                                                                                                    \\
    T_{\mu\nu}^{\text{Maxwell}}
    \equiv       & -\frac{1}{4}g_{\mu\nu}g^{\alpha\beta}g^{\rho\sigma}F_{\alpha\rho}F_{\beta\sigma} + g^{\alpha\beta}F_{\alpha\mu}F_{\beta\nu} - \frac{1}{2}g_{\mu\nu}m^2g^{\alpha\beta}A_{\alpha}A_{\beta} + m^2A_{\mu}A_{\nu} \;, \label{eq:MaxwellContribution}                                        \\
    T_{\mu\nu}^{\text{G}}
    \equiv       & \frac{1}{\xi} \bigg[ -\frac{1}{2}g_{\mu\nu}\left(g^{\alpha\beta}\nabla_\alpha A_\beta\right)^2 + \left(g_{\mu\nu}g^{\rho\sigma}A_\sigma\nabla_\rho - A_\nu\nabla_\mu - A_\mu\nabla_\nu\right)\left(g^{\alpha\beta}\nabla_\alpha A_\beta\right) \bigg] \;, \label{eq:GaugeContribution} \\
    T_{\mu\nu}^{\text{ghost}}
    \equiv       & ig_{\mu\nu}g^{\rho\sigma}\partial_{\rho}\bar{\chi}\partial_{\sigma}\chi - i\left(\partial_{\mu}\bar{\chi}\partial_{\nu}\chi + \partial_{\nu}\bar{\chi}\partial_{\mu}\chi\right) + ig_{\mu\nu}m_{\chi}^{2}\bar{\chi}\chi \label{eq:GhostContribution}\;.
  \end{align}
\end{subequations}
Following the procedure in Ref.~\cite{chu_adiabatic_2017}, one can define the components of the four-vector $A_\mu$ as a combination of temporal, transverse, and longitudinal parts,
\begin{equation}
  A_\mu \equiv \left(A_0,B_i+\partial_iA\right) \;.
\end{equation}
These components and the ghost field can be expanded in terms of mode functions $Y_a$, for $a=0,L,T,\chi$, where $L$ represents longitudinal and $T$ transverse contributions:
\begin{subequations}
  \begin{equation}
    A_0 = \frac{1}{m^2a^2}\int \frac{d^3\vec{k}}{\left(2\pi\right)^3} \left[a_{\vec{k}}^{(0)}\left(\partial_\eta - \frac{2a'}{a}\right)\left(maY_0\right)e^{i\vec{k}\cdot\vec{x}} + a_{\vec{k}}^{(3)}kmaY_Le^{i\vec{k}\cdot\vec{x}} + \text{H.c.}\right] \;,
  \end{equation}
  \begin{equation}
    B_i = \int\frac{d^3\vec{k}}{\left(2\pi\right)^3} \sum_{p=1,2}\left(\epsilon_i^{p}a_{\vec{k}}^{(p)}Y_Te^{i\vec{k}\cdot\vec{x}} + \text{H.c.}\right) \;,
  \end{equation}
  \begin{equation}
    A = \frac{1}{m^2a^2}\int\frac{d^3\vec{k}}{\left(2\pi\right)^3}\left(a_{\vec{k}}^{(0)}maY_0e^{i\vec{k}\cdot\vec{x}} - a_{\vec{k}}^{(3)}\partial_\eta\left(\frac{ma}{k}Y_L\right)e^{i\vec{k}\cdot\vec{x}} + \text{H.c.}\right) \;,
  \end{equation}
  \begin{equation}
    \chi = \int\frac{d^3\vec{k}}{\left(2\pi\right)^3}\left(\frac{b_{\vec{k}}Y_\chi}{a} e^{i\vec{k}\cdot\vec{x}} + \frac{b^\dagger_{\vec{k}}Y_\chi^*}{a} e^{-i\vec{k}\cdot\vec{x}} \right) \;,
  \end{equation}
  \begin{equation}
    \bar{\chi} = \int\frac{d^3\vec{k}}{\left(2\pi\right)^3}\left(\frac{\bar{b}_{\vec{k}}Y_\chi}{a} e^{i\vec{k}\cdot\vec{x}} + \frac{\bar{b}^\dagger_{\vec{k}}Y_\chi^*}{a} e^{-i\vec{k}\cdot\vec{x}} \right) \;,
  \end{equation}
\end{subequations}
where $a_{\vec{k}}^{(\mu)}$, $b_{\vec{k}}$, and $\bar{b}_{\vec{k}}$ and their hermitian conjugates are annihilation and creation operators, $\epsilon_i^p$ are the two polarization vectors of the transverse modes, and $\text{H.c.}$ represents the hermitian conjugate of all preceding terms.
The creation and annihilation operators satisfy the commutation and anticommutation relations
\begin{subequations}
  \begin{align}
    \left[a_{\vec{k}}^{(a)}, a_{\vec{k}'}^{(b)\dagger}\right] & = \eta^{ab}\left(2\pi\right)^3\delta^{(3)}(\vec{k}-\vec{k}') \;,                                                   \\
    \left\{b_{\vec{k}}, \bar{b}_{\vec{k}'}^\dagger\right\}    & = -\left\{\bar{b}_{\vec{k}}, b_{\vec{k}'}^\dagger\right\} = i\left(2\pi\right)^3\delta^{(3)}(\vec{k}-\vec{k}') \;,
  \end{align}
\end{subequations}
where $\eta^{ab}=\text{diag}(-1,1,1,1)$.
The polarization vectors satisfy
\begin{subequations}
  \begin{equation}
    \sum_{i}k_i\epsilon_i^{p} = 0 \;,
  \end{equation}
  \begin{equation}
    \sum_{i}\epsilon_i^{p}\epsilon_i^{p^\prime}=\delta^{pp^\prime} \;,
  \end{equation}
  \begin{equation}
    \sum_{p=1,2}\epsilon_i^p\epsilon_j^{p}=\delta_{ij}-\frac{k_ik_j}{k^2} \;,
  \end{equation}
\end{subequations}
with $p$ the polarization index.
The modes $Y_a$ satisfy the decoupled set of differential equation
\begin{equation}\label{eq:VectorModeDiffEq}
  \left(\partial_\eta^2 + \Omega_a^2\right) Y_{a} = 0 \;,
\end{equation}
where
%\begin{subequations}
%    \begin{align}
%        \Omega_0^2    & = k^2 + \xi m^2a^2 - \frac{a''}{a} \;,                         \\
%        \Omega_L^2    & = k^2 + m^2a^2 + \frac{a''}{a} - \frac{2a^{\prime 2}}{a^2} \;, \\
%        \Omega_T^2    & = k^2 + m^2a^2 \;,                                             \\
%        \Omega_\chi^2 & = k^2 + m_\chi^2a^2 - \frac{a''}{a} \;.
%    \end{align}
%\end{subequations}
\begin{equation}
  \Omega_a^2 \equiv \omega_a^2 + \zeta_a
\end{equation}
is defined for each contribution by
\begin{subequations}
  \begin{align}
    \omega_0^2     & \equiv k^2 + \xi m^2a^2 \;, \\
    \omega_{L,T}^2 & \equiv k^2 + m^2a^2 \;,     \\
    \omega_\chi^2  & \equiv k^2+m_\chi^2a^2 \;,
  \end{align}
\end{subequations}
and
\begin{subequations}
  \begin{align}
    \zeta_{0,\chi} & \equiv -\frac{a''}{a} \;,                            \\
    \zeta_{L}      & \equiv \frac{a''}{a} - \frac{2a^{\prime 2}}{a^2} \;, \\
    \zeta_{T}      & \equiv 0 \;.
  \end{align}
\end{subequations}
The modes satisfy the standard normalization conditions for a scalar field,
\begin{equation}\label{eq:VectorNormalizationY}
  Y_a Y_a^{*\prime} - Y_a' Y_a^* = i \;.
\end{equation}
Note that in the massless limit $Y_0$ and $Y_\chi$ satisfy the same differential equations, so we can choose
\begin{equation}\label{eq:VectorMasslessModeEquivalence}
  Y_0 = Y_\chi \quad \text{for} \quad m=m_\chi=0 \;.
\end{equation}

Using this mode decomposition, one can find contributions to the energy density $\rho\equiv\langle T_{\hat{0}\hat{0}}\rangle$ from Eqs.~\eqref{eq:MaxwellContribution}--\eqref{eq:GhostContribution} in terms of $Y_a$:
\begin{subequations}
  \begin{align}
    \rho^{\text{Maxwell}}
    = & \masslesslimit\frac{1}{4\pi^2a^4}\int_0^\Lambda dk\,k^2 \bigg[ \omega_T^2\magsq{Y_T} + \magsq{Y_T^\prime} - \left(k^2 + \frac{a^{\prime 2}}{a^2}\right)\magsq{Y_0} + \frac{a'}{a}\partial_\eta\left(\magsq{Y_0}\right) - \magsq{Y_0^{\prime}} \nonumber                                  \\
      & + \left( \omega_L^2 + \frac{a^{\prime 2}}{a^2} \right)\magsq{Y_L} + \frac{a'}{a}\partial_\eta\left(\magsq{Y_L}\right) + \magsq{Y_L^\prime} \bigg] \label{eq:VectorEnergyDensityMaxwellContribution}                                                                                      \\
    \rho^{\text{G}}
    = & \masslesslimit\frac{1}{2\pi^2a^4}\int_0^\Lambda dk\,k^2 \left[ \left(\omega_0^2 - \frac{1}{2}\xi m^2a^2 + \frac{a^{\prime 2}}{a^2}\right)\magsq{Y_0} - \frac{a'}{a}\partial_\eta\left(\magsq{Y_0}\right) + \magsq{Y_0^\prime} \right] \label{eq:VectorEnergyDensityGaugeContribution}    \\
    \rho^{\text{ghost}}
    = & \lim\limits_{m_\chi\to 0}\frac{1}{2\pi^2a^4}\int_0^\Lambda dk\,k^2 \left[ -\left(\omega_\chi^2 + \frac{a^{\prime 2}}{a^2}\right)\magsq{Y_\chi} + \frac{a'}{a}\partial_\eta\left(\magsq{Y_\chi}\right) - \magsq{Y_\chi^\prime} \right] \;,\label{eq:VectorEnergyDensityGhostContribution}
  \end{align}
\end{subequations}
where $\Lambda$ is a cutoff regulator which we will later demonstrate can be taken to infinity in the massless limit.

One may adiabatically renormalize the energy density by writing the vacuum states with the standard WKB ansatz
\begin{align}\label{eq:VectorWkbAnsatz}
  Y_a = \frac{1}{\sqrt{2W_a}}e^{-i\int^\eta d\bar{\eta} W_a(\bar{\eta})} \;,
\end{align}
where
\begin{equation}\label{eq:VectorWa}
  \left(W_a\right)^2 = \Omega_a^2 - \left[\frac{W_a''}{2W_a} - \frac{3}{4}\left(\frac{W_a'}{W_a}\right)^2\right] \;.
\end{equation}
Solutions to Eq.~\eqref{eq:VectorWa} can be approximated using $W_a^{(0)} = \Omega_a$ as the lowest order and iterating to higher orders, keeping to the appropriate adiabatic order, given by the number of time derivatives on the scale factor, at each iteration.
Substitution of $W_a$ to some adiabatic order $A$ into Eq.~\eqref{eq:VectorWkbAnsatz} would then require expanding the square root only to terms of adiabatic order $A$.

In order to renormalize the energy density, one must take Eq.~\eqref{eq:VectorWkbAnsatz} to the appropriate adiabatic order and substitute into Eqs.~\eqref{eq:VectorEnergyDensityMaxwellContribution}--\eqref{eq:VectorEnergyDensityGhostContribution} to produce the renormalization counterterms.
On dimensional grounds, one would need to keep to fourth adiabatic order to renormalize $\rho$.
The fourth order counterterms are
\begin{align}
  \rho_c^{(4)} = & \masslesslimit \frac{1}{4a^2} \int \frac{d^3k}{(2\pi)^3}\bigg\{ W_0 + W_L + 2W_T - 2W_\chi + \frac{\omega_0^2}{W_0} + \frac{\omega_L^2}{W_L} + \frac{2\omega_T^2}{W_T} - \frac{2\omega_\chi^2}{W_\chi} + \frac{a^{\prime 2}}{a^2W_0}\nonumber                                                                                       \\
                 & + \frac{a^{\prime 2}}{a^2W_L} - \frac{2a^{\prime 2}}{a^2W_\chi} + \frac{a'W_0'}{aW_0^2} - \frac{a'W_L'}{aW_L^2} - \frac{2a'W_\chi'}{aW_\chi^2} + \frac{W_0^{\prime 2}}{4W_0^3} + \frac{W_L^{\prime 2}}{4W_L^3} + \frac{W_T^{\prime 2}}{2W_T^3} - \frac{W_\chi^{\prime 2}}{2W_\chi^3} \bigg\}^{(4)} \;,\label{eq:VectorCounterterms}
\end{align}
where $\{\dots\}^{(4)}$ implies that all $W_a$ are taken to fourth order.

In order to analyze the early-time behavior of the energy density, we will parameterize the mode functions in terms of zeroth-order adiabatic states
\begin{equation}\label{eq:VectorApproxModeParameterization}
  Y_a = \alpha_{k,a} Y_a^{(0)} + \beta_{k,a} Y_a^{(0)*} \;,
\end{equation}
\begin{equation}\label{eq:VectorApproxModePrimeParameterization}
  Y_a' = \alpha_{k,a} Y_a^{(0)\prime} + \beta_{k,a} Y_a^{(0)*\prime} \;.
\end{equation}
From the normalization condition~\eqref{eq:VectorNormalizationY}, one has
\begin{equation}
  \magsq{\alpha_{k,a}}-\magsq{\beta_{k,a}} = 1 \;.
\end{equation}
One could instead parameterize the mode functions in terms of higher-order adiabatic states, in which case the Bogoliubov coefficients $\alpha_{k,a}$ and $\beta_{k,a}$ would be constant to the given adiabatic order, but zeroth order will be sufficient to properly renormalize the theory.
Substituting Eqs.~\eqref{eq:VectorApproxModeParameterization} and \eqref{eq:VectorApproxModePrimeParameterization} into Eq.~\eqref{eq:VectorModeDiffEq},
one obtains differential equations for the coefficients,
\begin{subequations}
  \begin{align}
    \alpha_{k,a}' & = \frac{\Omega_a'}{2\Omega_a}\beta_{k,a}e^{2i\theta_a}       \\
    \beta_{k,a}'  & = \frac{\Omega_a'}{2\Omega_a}\alpha_{k,a}e^{-2i\theta_a} \;,
  \end{align}
\end{subequations}
where $\theta_a \equiv \int^\eta d\bar{\eta}W_a(\bar{\eta})$.

One can then obtain the renormalized energy density in terms of $\alpha_{k,a}$ and $\beta_{k,a}$ by subtracting these renormalization counterterms from the unrenormalized energy density given in Eqs.~\eqref{eq:VectorEnergyDensityMaxwellContribution}--\eqref{eq:VectorEnergyDensityGhostContribution}.
Using a zeroth order parameterization, one finds that the renormalized energy density $\rho_r$ separates into analytic and mode terms
\begin{equation}
  \rho_r = \rho_{\text{an}} + \rho_{\alpha\beta} \;,
\end{equation}
where the analytic terms $\rho_{\text{an}}$ are finite higher-order terms, independent of the cutoff regulator $\Lambda$, coming from the counterterms~\eqref{eq:VectorCounterterms},
\begin{align}
  \rho_{\text{an}}
  = & \frac{1}{2880\pi^2}\left[ 62\,{}^{(3)}H_{00} + \left(3 + \frac{5}{2}\ln\xi\right)\,{}^{(1)}H_{00} \right] \nonumber                                         \\
  = & \frac{1}{2880\pi^2}\bigg[ \frac{186a^{\prime 4}}{a^6} + \frac{216a^{\prime 2}a''}{a^5} + \frac{54a^{\prime\prime 2}}{a^4} - \frac{108a'a'''}{a^4} \nonumber \\
    & + \ln\xi\left( \frac{180a^{\prime 2}a''}{a^5} + \frac{45a^{\prime\prime 2}}{a^4} - \frac{90a'a'''}{a^4}\right) \bigg] \;,
\end{align}
where ${}^{(1)}H_{00}$ and ${}^{(3)}H_{00}$ are higher order corrections to the Einstein field equations~\cite{birrell_quantum_1982}, and the mode terms $\rho_{\alpha\beta}$ are those coming from the zeroth order parameterization of Eq.~\eqref{eq:VectorApproxModeParameterization},
\begin{align}
  \rho_{\alpha\beta}
  \equiv & \lim\limits_{m,m_\chi\to 0} \frac{1}{a^2}\int_0^\Lambda \frac{d^3k}{(2\pi)^3}\bigg[
  % magsq
  \frac{k^2+\omega_T^2}{a^2\omega_T}\magsq{\beta_{k,T}}
  + \frac{m^2}{2\omega_L}\magsq{\beta_{k,L}}
  + \frac{1}{\omega_0}\left(\frac{3\omega_0^2+k^2}{2a^2}+\frac{a^{\prime 2}}{a^4}\right)\magsq{\beta_{k,0}} \nonumber                             \\
         & - \frac{1}{\omega_\chi}\left( \frac{\omega_\chi^2 + k^2}{a^2} + \frac{a^{\prime 2}}{a^4}\right)\magsq{\beta_{k,\chi}}
  % Re
  - \frac{m^2}{\omega_T}\Re{\alpha_{k,T}\beta_{k,T}^*e^{-2i\theta_T}}
  + \frac{m^2}{2\omega_L}\Re{\alpha_{k,L}\beta_{k,L}^*e^{-2i\theta_L}} \nonumber                                                                  \\
         & + \frac{1}{\omega_0}\left(\frac{a^{\prime 4}}{a^2}-\frac{\xi m^2}{2a^2}\right)\Re{\alpha_{k,0}\beta_{k,0}^*e^{-2i\theta_0}}
  + \frac{1}{\omega_\chi}\left(m_\chi^2 -\frac{a^{\prime 2}}{a^4}\right)\Re{\alpha_{k,\chi}\beta_{k,\chi}^*e^{-2i\theta_\chi}} \nonumber          \\
  % Im
         & - \frac{2a'}{a^3}\Im{\alpha_{k,0}\beta_{k,0}^*e^{-2i\theta_0}} + \frac{2a'}{a^3}\Im{\alpha_{k,\chi}\beta_{k,\chi}^*e^{-2i\theta_\chi}}
  \bigg] \;. \label{eq:VectorRenormalizedEnergyDensityWithIntegral}
\end{align}

Assuming $\beta_{k,a}$ falls faster than $k^{-2}$, integrating the terms in Eq.~\eqref{eq:VectorRenormalizedEnergyDensityWithIntegral} will yield finite results, even if $\Lambda\to\infty$, and hence the massless limit can be freely taken inside the integral.
One finds from substitution of Eqs.~\eqref{eq:VectorWkbAnsatz}, \eqref{eq:VectorApproxModeParameterization}, and \eqref{eq:VectorApproxModePrimeParameterization} into Eq.~\eqref{eq:VectorModeDiffEq} that $\alpha_{k,0}'=\alpha_{k,\chi}'$ and $\beta_{k,0}'=\beta_{k,\chi}'$ after taking the massless limit, which when combined with Eq.~\eqref{eq:VectorMasslessModeEquivalence} allows one to choose the coefficients for the $0$ and $\chi$ contributions to be identical.
The mode term contribution to the energy density therefore drastically simplifies to
\begin{equation}\label{eq:VectorRenormalizedEnergyDensity}
  \rho_{\alpha\beta} = \frac{1}{\pi^2a^4}\int_0^{\Lambda}dk\,k^3\magsq{\beta_{k,T}} \;.
\end{equation}
The ultraviolet cutoff can now be removed, and we take the limit $\Lambda\to\infty$.
%One also finds in the massless limits that the coefficients for the $T$ contribution are constant,
%\begin{equation}
%    \beta_{k,T}' = 0 \;,
%\end{equation}
%so one can safely take the cutoff $\Lambda\rightarrow\infty$ inside the integral in Eq.~\eqref{eq:VectorRenormalizedEnergyDensity} without introducing any divergences.
The renormalized energy density ultimately only depends on the transverse mode functions, which are the only physical modes of the theory, and higher order dependencies on the background curvature:
\begin{align}\label{eq:VectorFinalRenormalizedEnergyDensity}
  \rho_r = & \frac{1}{\pi^2a^4}\int_0^\infty dk\; k^3\magsq{\beta_{k,T}} + \frac{1}{2880\pi^2}\left[ 62\,{}^{(3)}H_{00} + \left(3 + \frac{5}{2}\ln\xi\right)\,{}^{(1)}H_{00} \right] \;.
\end{align}
Assuming the higher order corrections are subdominant in the semiclassical approximation below the Planck scale, $\rho_r$ for massless vector fields does have the expected radiation-dominated behavior.
This result agrees with the prediction in~\cite{wald_1978} that other than the anomalous term all of the contributions in a flat FLRW metric of a conformally invariant field will act like radiation.

The higher-order terms in~\eqref{eq:VectorFinalRenormalizedEnergyDensity} are of the same form as those found for the trace anomaly in~\cite{endo_1984,chu_adiabatic_2017}, in which the ${}^{(1)}H_{00}$ term has a gauge-dependent coefficient.
This coefficient corresponds to a $\Box R$ term appearing in the trace anomaly.
The exact value of this coefficient is dependent on the regularization scheme used, unlike for scalar and spin-$\frac{1}{2}$ fields which respectively have the same coefficient regardless of regularization scheme.

\section{Energy Density for Dirac Fields\label{sec:SpinorFields}}
We now turn our attention to Dirac spinor fields in the preinflationary era of an FLRW universe.
We follow the work and notation given in Ref.~\cite{rio_renormalized_2014}.
There, expressions are in terms of cosmic time $t$, related to conformal time by $a\, d\eta = dt$.
We summarize the procedure for obtaining the unrenormalized energy density here.
%in which spin-half fields are considered using cosmic time $t$.
%Because we would instead like to work with conformal time $\eta$, defined by $a\,d\eta=dt$, we summarize the procedure for obtaining the unrenormalized energy density here.

Consider Dirac spinor fields $\Psi(x)$ that obey the Dirac equation in curved spacetime,
\begin{equation}
  \left( i\gamma^{a}e\indices{_a^\mu}\nabla_{\mu} - m \right) \Psi = 0 \;,\label{eq:DiracEq}
\end{equation}
where $e\indices{_a^\mu} $ is the vierbein, $\gamma^{a}$ are the flat spacetime Dirac matrices satisfying $\{\gamma^{a},\gamma^{b}\}=2\eta^{ab}$, and $\nabla_\mu\equiv\partial_\mu+\Gamma_\mu$ is the covariant derivative associated with the spin connection $\Gamma_\mu$.
For the metric~\eqref{eq:FLRWMetric}, the Dirac equation~\eqref{eq:DiracEq} becomes
\begin{equation}
  \left[ \gamma^0\left(\partial_\eta + \frac{3a'}{2a}\right) + \gamma^{i}\partial_i + ima \right] \Psi = 0 \;.
\end{equation}
The field can be written in terms of creation and annihilation operators $D^{\dagger}_{\vk\lambda}(\eta)$ and $B_{\vk\lambda}(\eta)$ as
\begin{equation}
  \Psi = \sum_{\lambda=\pm\frac{1}{2}} \int d^{3}k \left(B_{\vk\lambda}\psi_{\vk\lambda} + D^{\dagger}_{\vk\lambda}C\bar{\psi}_{\vk\lambda}^\intercal \right) \;,
\end{equation}
where $C$ is the charge conjugation matrix, $\lambda=\pm1/2$ represents the helicity eigenvalue, and
\begin{equation}
  \left\{B_{\vec{k},\lambda}, B_{\vec{k},\lambda}^\dagger\right\} = \delta_{\lambda\lambda'}\delta^{(3)}(\vec{k}-\vec{k}')
\end{equation}
and similarly for $D_{\vec{k}\lambda}$ and $D_{\vec{k}\lambda}^\dagger$, with all other anticommutators vanishing.
Working in the Dirac-Pauli representation for $\gamma^{a}$, the modes $\psi_{\vk\lambda}$ can be written as
\begin{equation}
  \psi_{\vk\lambda}(\eta,\vec{x}) = \frac{e^{i\vk\cdot\vec{x}}}{\sqrt{8\pi^3 a^3}}
  \begin{pmatrix}
    h^{I}_{k}(\eta)\xi_{\lambda}(\vk) \;, \\
    h^{II}_{k}(\eta)\hat{k}\cdot\vec{\sigma}\xi_{\lambda}(\vk)
  \end{pmatrix} \;,
\end{equation}
where $\xi_{\lambda}(\vk)$ are two-component spinors and are eigenvectors of the spin component along the $\vk$ direction, so that $\frac{1}{2}(\hat{k}\cdot\vec{\sigma})\xi_{\lambda}(\vk) = \lambda\xi_{\lambda}(\vk)$, with normalization $\xi^{\dagger}_{\lambda}\xi_{\lambda}=1$, and $h^{I}_k$ and $h^{II}_k$ are scalar functions that satisfy coupled first-order differential equations
\begin{subequations}
  \begin{align}
    \partial_\eta h^{I}_k(\eta)  & = -ikh^{II}_k(\eta) - ima(\eta)h^{I}_k(\eta) \;, \label{eq:TwoSpinorDiffEqI}  \\
    \partial_\eta h^{II}_k(\eta) & = -ikh^{I}_k(\eta) + ima(\eta)h^{II}_k(\eta) \label{eq:TwoSpinorDiffEqII} \;,
  \end{align}
\end{subequations}
and have the normalization
\begin{equation}\label{eq:TwoSpinorNormalization}
  \left| h^{I}_k(\eta) \right|^2 + \left| h^{II}_k(\eta) \right|^2 = 1 \;.
\end{equation}
The energy density for the Dirac field in terms of the mode functions $h_k^{I,II}$ can be written as
\begin{equation}\label{eq:SpinorEnergyDensityModeFunctions}
  \rho = \frac{1}{\pi^2a^4} \int_0^\infty dk k^2 \left[ma \left( \left| h^{II}_k \right|^2 - \left| h^{I}_k \right|^2 \right) - k \left( h^{I}_{k}h^{II*}_k + h^{I*}_kh^{II}_k \right) \right] \;.
\end{equation}

At this point, we will diverge from this procedure coming from~\cite{rio_renormalized_2014}, who themselves proceeded to obtain counterterms to a generic unrenormalized energy in an FLRW universe.
These counterterms were then used to prove conservation of the energy density and were applied to a de Sitter spacetime and a radiation-dominated universe.
As demonstrated there, one can make use of adiabatic regularization to renormalize the energy density, but the ansatz used for the WKB approximation of adiabatic states must be a modified form of the standard ansatz.
This renormalization procedure was first demonstrated in~\cite{landete_adiabatic_2013} and has been applied in other cases~\cite{landete_adiabatic_2014, torrenti_adiabatic_2015, barbero_adiabatic_2018, palau_adiabatic_2020}.
Here, we will use the energy density~\eqref{eq:SpinorEnergyDensityModeFunctions} as well as the modified WKB ansatz to obtain renormalization counterterms, but we will use a different process to obtain an explicit form of the unrenormalized energy density.
Namely, we will expand the mode functions $h_k^{I,II}$ in terms of adiabatic modes $g_k^{I,II}$ via a Bogoliubov-like expansion in order to analyze early time behavior using the Bogoliubov coefficients.

First, one expands $h_k^{I,II}$ in terms of adiabatic modes $g^{I,II}_k$,
\begin{subequations}
  \begin{align}
    h^{I}_k = \alpha_k g^{I}_k - \beta_k g_k^{II*} \;, \label{eq:TwoSpinorAdiabaticFormI} \\
    h^{II}_k = \alpha_k g^{II}_k + \beta_k g_k^{I*} \label{eq:TwoSpinorAdiabaticFormII}\;,
  \end{align}
\end{subequations}
where $g^{I,II}_k$ are given to adiabatic order $A$,
\begin{equation}
  g^{I,II}_k = g^{I,II(0)}_k + g^{I,II(1)}_k + \dots + g^{I,II(A)}_k \;,
\end{equation}
with adiabatic order understood to be the number of conformal time derivatives,
and satisfy the differential equations~\eqref{eq:TwoSpinorDiffEqI} and \eqref{eq:TwoSpinorDiffEqII}, and $\alpha_k$ and $\beta_k$ are
the time-dependent Bogoliubov coefficients and are constant to order $A$.
Coupled first order differential equations for $\alpha_k$ and $\beta_k$ can be obtained from Eqs.~\eqref{eq:TwoSpinorDiffEqI}, \eqref{eq:TwoSpinorDiffEqII}, and \eqref{eq:TwoSpinorNormalization}.
The unrenormalized energy density in terms of the adiabatic states is
\begin{align}\label{eq:AdiabaticStatesEnergyDensity}
  \rho = & \frac{1}{\pi^2a^4}\int_0^\infty dkk^2 \bigg\{ 2\left|\beta_k\right|^2 \left[ ma\left(\left|\gI\right|^2 - \left|\gII\right|^2\right) + 2k\text{Re}\left(\gI\gIIc\right) \right] + 4ma\text{Re}\left( \alpha_k\beta_k^*\gI\gII \right) \nonumber \\
         & + 2k\text{Re}\left[ \alpha_k\beta_k^*\left( \left(\gII\right)^2 - \left(\gI\right)^2 \right) \right] + ma\left(\left|\gII\right|^2 - \left|\gI\right|^2\right) - 2k\text{Re}\left(\gI\gIIc\right) \bigg\} \;,
\end{align}
and the adiabatic renormalization counterterms are
\begin{align}\label{eq:AdiabaticStatesEnergyDensityCounterterms}
  \rho_c = & \frac{1}{\pi^2a^4}\int_0^\infty dkk^2 \left[ ma\left(\left|\gII\right|^2 - \left|\gI\right|^2\right) - 2k\text{Re}\left(\gI\gIIc\right) \right]^{(4)} \;,
\end{align}
where $[\dots]^{(4)}$ indicates that $\gI$, $\gII$ are fourth order states.
In order for the energy density to be renormalized and all divergences eliminated, the adiabatic states in Eq.~\eqref{eq:AdiabaticStatesEnergyDensity} must be at least of the adiabatic order at which the counterterms in Eq.~\eqref{eq:AdiabaticStatesEnergyDensityCounterterms} are divergent.

In order to obtain forms for the adiabatic states, one can obtain uncoupled second order equations from Eqs.~\eqref{eq:TwoSpinorDiffEqI} and \eqref{eq:TwoSpinorDiffEqII},
\begin{align}
   & \left( \partial_\eta^2 + \frac{a'}{a}\partial_\eta - ima' + \omega^2 \right) g^{I}_k = 0 \;,  \\
   & \left( \partial_\eta^2 + \frac{a'}{a}\partial_\eta + ima' + \omega^2 \right) g^{II}_k = 0 \;,
\end{align}
where
\begin{equation}
  \omega^2 \equiv k^2 + m^2a^2 \label{eq:Omega}\;,
\end{equation}
and assume formal WKB series solutions, truncating at the desired adiabatic order.
However, Ref.~\cite{landete_adiabatic_2013} and later Refs.~\cite{landete_adiabatic_2014,rio_renormalized_2014} pointed out that the usual WKB ansatz does not satisfy the normalization condition~\eqref{eq:TwoSpinorNormalization}, so one must use a modified WKB ansatz of the form
\begin{subequations}
  \begin{align}
    g^{I}_k  & = \sqrt{\frac{\omega+ma}{2\omega}} F e^{-i\theta_k} \label{eq:AdiabaticStatesWKBI} \;, \\
    g^{II}_k & = \sqrt{\frac{\omega-ma}{2\omega}} G e^{-i\theta_k} \label{eq:AdiabaticStatesWKBII}\;,
  \end{align}
\end{subequations}
and the functions
\begin{subequations}
  \begin{align}
    F        & = 1 + F^{(1)} + \dots + F^{(A)} \;,                                                  \\
    G        & = 1 + G^{(1)} + \dots + G^{(A)} \;,                                                  \\
    \theta_k & = \int^\eta d\tilde{\eta} \left(\omega + \omega^{(1)} + \dots + \omega^{(A)} \right)
  \end{align}
\end{subequations}
are determined by repeated substitution of Eqs.~\eqref{eq:AdiabaticStatesWKBI} and \eqref{eq:AdiabaticStatesWKBII} into Eqs.~\eqref{eq:TwoSpinorDiffEqI}, \eqref{eq:TwoSpinorDiffEqII}, and \eqref{eq:TwoSpinorNormalization}.
There is an ambiguity in the exact forms following this method, but all local observables are independent of the ambiguity~\cite{landete_adiabatic_2013}, so one may fix the ambiguity by choosing $F^{(n)}(-m)=G^{(n)}(m)$ for each order $n\ge 1$.
However, we are only interested in using zeroth order states, for which one obtains
\begin{subequations}
  \begin{align}
    \gI  & = \sqrt{\frac{\omega+ma}{2\omega}}e^{-i\theta_k} \label{eq:AdiabaticStatesZerothOrderI} \;, \\
    \gII & = \sqrt{\frac{\omega-ma}{2\omega}}e^{-i\theta_k} \label{eq:AdiabaticStatesZerothOrderII}\;,
  \end{align}
\end{subequations}
which have a normalization from Eq.~\eqref{eq:TwoSpinorNormalization} of
\begin{equation}
  \magsq{\gI} + \magsq{\gII} = 1 \;.
\end{equation}
Note that we will continue writing $\theta_k$ like we have in Eqs.~\eqref{eq:AdiabaticStatesZerothOrderI} and \eqref{eq:AdiabaticStatesZerothOrderII} for simplicity and assume it to be understood that only the zeroth order term is kept.
Substituting Eqs.~\eqref{eq:TwoSpinorAdiabaticFormI} and \eqref{eq:TwoSpinorAdiabaticFormII} into the differential equations~\eqref{eq:TwoSpinorDiffEqI} and \eqref{eq:TwoSpinorDiffEqII}, one obtains differential equations for $\alpha_k$ and $\beta_k$,
\begin{align}
  \alpha_k' & = \frac{-kma'}{2\omega^2}\beta_k e^{2i\theta_k} \label{eq:AlphaDiffEq0thOrder} \;, \\
  \beta_k'  & = \frac{kma'}{2\omega^2}\alpha_k e^{-2i\theta_k} \label{eq:BetaDiffEq0thOrder}\;,
\end{align}
and substituting them into the normalization condition~\eqref{eq:TwoSpinorNormalization}, one finds
\begin{equation}\label{eq:AlphaBetaNormalization}
  \magsq{\alpha_k} + \magsq{\beta_k} = 1 \;.
\end{equation}

\subsection{Renormalized Energy Density\label{sec:SpinorFieldsRenormalizedEnergyDensity}}
In order to obtain finite results so that we may inspect the behavior of the energy density, one must subtract counterterms up to fourth order from Eq.~\eqref{eq:AdiabaticStatesEnergyDensity}.
Order by order, these counterterms $\rho^{(n)}$ are
\begin{align}
  \rho^{(0)}_c = & -\omega \label{eq:0thOrderCounterterms} \;,           \\
  \rho^{(2)}_c = & \frac{k^2\omega^{\prime 2}}{8m^2a^2\omega^3} \;,
  = \frac{k^2m^2a^{\prime 2}}{8\omega^5} \label{eq:2ndOrderCounterterms} \\
  \rho^{(4)}_c = & \order{k^{-5}} \label{eq:4thOrderCounterterms}\;.
\end{align}
The fourth order counterterms produce finite contributions to the energy density.
Using the zeroth order adiabatic states~\eqref{eq:AdiabaticStatesZerothOrderI} and \eqref{eq:AdiabaticStatesZerothOrderII} in Eq.~\eqref{eq:SpinorEnergyDensityModeFunctions} and subtracting the counterterms~\eqref{eq:0thOrderCounterterms}-\eqref{eq:4thOrderCounterterms}, one obtains the renormalized energy density
\begin{equation}
  \rho_r
  = \frac{1}{\pi^2a^4} \int_0^\infty dk\, k^2 \left[ 2\omega \magsq{\beta_k} - \frac{k^2m^2a^{\prime 2}}{8\omega^5} \right] + \frac{2}{2880\pi^2}\left[ -\frac{1}{2}\tensor*[^{(1)}]{H}{_{00}} + \frac{11}{2}\tensor*[^{(3)}]{H}{_{00}} \right] \label{eq:RenormalizedEnergyDensity0thOrder} \;,
\end{equation}
where the finite renormalization terms coming from the fourth order counterterms~\cite{birrell_quantum_1982},
\begin{align}
  \frac{2}{2880\pi^2}\left[-\frac{1}{2}\tensor*[^{(1)}]{H}{_{00}} + \frac{11}{2}\tensor*[^{(3)}]{H}{_{00}} \right]
   & = \frac{2}{2880\pi^2a^4}\left( \frac{33a'^4}{2a^4} + \frac{18a'a'''}{a^2} - \frac{9a''^2}{a^2} - \frac{36a'^2a''}{a^3} \right) \;, \label{eq:FiniteRenormalizationCounterterms}
\end{align}
are assumed small beyond the Planck era.

We ultimately will attempt to solve the Friedmann equation
\begin{equation}\label{eq:Friedmann}
  H^2 = \frac{8\pi}{3}\rho \;,
\end{equation}
where
\begin{equation}\label{eq:HubbleParameter}
  H \equiv \frac{a'}{a^2} \;,
\end{equation}
and $\rho$ contains $\rho_r$ and may also include other terms such as a cosmological constant or other classical contributions.
Because we are working with a semiclassical approximation, we do not assume our analysis to be valid during the Planck era.
Hence we will work starting at an initial time $\eta_0$, which is assumed to be after the Planck era and corresponds to a scale factor $a_0$ that is above the Planck scale, and demonstrate that $\rho_r\sim a^{-4}$ for some region $\eta_0<\eta<\eta_1$.

Above the Planck scale, the Hubble parameter $H\lesssim 1$, so given Eqs.~\eqref{eq:HubbleParameter} and \eqref{eq:FiniteRenormalizationCounterterms}, we will insist that the following set of inequalities of derivatives of the scale factor must hold:
\begin{align}
  a'     & \lesssim a^2 \label{eq:ScaleFactorPrime}\;,  \\
  a''    & \lesssim a^3 \label{eq:ScaleFactorPrime2}\;, \\
  a'a''' & \lesssim a^6 \label{eq:ScaleFactorPrime3}\;.
\end{align}
We will use these inequalities in order to compute the integral in Eq.~\eqref{eq:RenormalizedEnergyDensity0thOrder}.

\subsection{Bounds on the Renormalized Energy Density\label{sec:SpinorFieldsBounds}}
We will begin by splitting the integral into infrared and ultraviolet regions $I_1$ and $I_2$ by a cutoff $k_c$.
The infrared contribution is
\begin{align}
  I_1 & = \int_0^{k_c} dk\,k^2\left[ 2\omega \magsq{\beta_k} - \frac{k^2m^2a^{\prime 2}}{8\omega^5} \right] \nonumber                                           \\
      & = \int_{0}^{k_c}dk\,k^2\left[ 2k\magsq{\beta_k} + 2\left(\omega-k\right)\magsq{\beta_k} - \frac{k^2m^2a^{\prime 2}}{8\omega^5} \right] \;,\label{eq:I1}
\end{align}
and the ultraviolet contribution is
\begin{align}\label{eq:I2}
  I_2 & = \int_{k_c}^\infty dk\,k^2\left[ 2\omega\magsq{\beta_k} - \frac{k^2m^2a^{\prime 2}}{8\omega^5} \right] \;.
\end{align}
At time $\eta_0$, we define
\begin{equation}\label{eq:B0}
  B_0\equiv \int_0^{k_c}dk\,2k^3\magsq{\beta_k(\eta_0)} \;.
\end{equation}
If $B_0$ is the dominant contribution to the renormalized energy density at time $\eta$, then $\rho_r\propto a^{-4}$ as desired.
However, as one may anticipate given the apparent logarithmically divergent term in Eq.~\eqref{eq:RenormalizedEnergyDensity0thOrder}, this may not always be the case.
We will investigate this in the following sections.
Given that $\abs{\beta_k(\eta_0)}\le 1$ from Eq.~\eqref{eq:AlphaBetaNormalization}, one finds from Eq.~\eqref{eq:B0} a lower bound on $k_c$ of
\begin{equation}\label{eq:kcLowerBound}
  k_c \gtrsim B_0^{1/4} \;.
\end{equation}

\subsubsection{Ultraviolet Region\label{sec:UltravioletRegion}}
One is tempted to assume $\beta_k\rightarrow 0$ sufficiently quickly at high $k$, as is often done with scalar fields~\cite{anderson_semiclassical_2020}.
However, because the final term in Eq.~\eqref{eq:I2} produces a logarithm divergence, it is evident that doing so will introduce a divergent contribution to the energy density.
This situation is occuring because until now we have been working with a zeroth order parameterization of the states, but because the logarithmic divergence comes in at higher than zeroth adiabatic order, one would expect to need to work with at least a second order parameterization to eliminate this higher-order divergence.
These problems would indeed disappear working with a higher-order parameterization, but it becomes much more difficult to analytically obtain generic bounds on $I_2$ doing so.

One can instead continue to work with a zeroth-order parameterization using $\alpha_k$ and $\beta_k$.
We can understand what the appropriate higher order states look like at large $k$ by integrating Eq.~\eqref{eq:BetaDiffEq0thOrder} by parts, which shows that the asymptotic behavior of $\beta_k$ will take the form
\begin{equation}
  \beta_k\xrightarrow[k\to\infty]{}-\frac{ikma'}{4\omega^3}\alpha_ke^{-2i\theta_k} \;.
\end{equation}
This motivates us to use the parameterization $\bar{\beta}_k$, defined as
\begin{equation}\label{eq:BetaBar}
  \bar{\beta_k} \equiv \beta_k + \frac{ikma'}{4\omega^3} \alpha_k e^{-2i\theta_k} \;,
\end{equation}
as the appropriate description to use in calculating the energy density at large $k$.
We then anticipate that $\bar{\beta}_k$ will fall faster than the leading order term at large $k$, so
\begin{equation}
  \abs{\bar{\beta}_k(\eta_0)} < \frac{A_0}{k^2}\left(\frac{k_c}{k}\right)^{b_0} \label{eq:BetaBarBound}
\end{equation}
for $k>k_c$, some $b_0>0$, and $A_0$ independent of $k$.
The ultraviolet integral~\eqref{eq:I2} written in terms of $\bar{\beta}_k$ is
\begin{align}
  I_2 & =
  %	\int_{k_c}^{\infty} dk\,k^2\left[ 2\omega\magsq{\bar{\beta}_k} - \frac{kma'}{2\omega^2}\Im{\alpha_k\bar{\beta}_k^*e^{-2i\theta_k}} - \frac{k^2m^2a^{\prime 2}}{8\omega^5}\magsq{\beta_k} \right] \nonumber                                                                  \\
  %	    & = 
  \int_{k_c}^\infty dk\,k^2\left[ \left( 1-\frac{k^2m^2a^{\prime 2}}{16\omega^6} \right) \left( 2\omega\magsq{\bar{\beta}_k} + \frac{kma'}{\omega^2}\Im{\alpha_k\bar{\beta}_k^* e^{-2i\theta_k}} \right) - \frac{k^4m^4a^{\prime 4}}{128\omega^{11}}\magsq{\alpha_k} \right]
  \;.\label{eq:I2BetaBar}
\end{align}
Because $\bar{\beta}_k$ encodes the cancellation of the divergent term in the renormalized energy density, one expects that the contributions to the energy density from Eq.~\eqref{eq:I2BetaBar} will converge.

In order for the energy density to be radiation-dominated the ultraviolet contribution~\eqref{eq:I2BetaBar} must either be the dominant contribution and itself radiation-dominated or be subdominant to the radiation-dominated part of the infrared contribution~\eqref{eq:I1}.
%must be either radiation-dominated or subdominant to that from the infrared region~\eqref{eq:I1}, assuming the infrared contribution is itself radiation-dominated.
As we will demonstrate, every term in Eq.~\eqref{eq:I2BetaBar} is in fact subdominant to $B_0$~\eqref{eq:B0}, which itself produces a radiation-dominated term in the energy density, and hence the latter is true.
%One can see from Eq.~\eqref{eq:BetaBar} that $\bar{\beta}_k$ will have a similar contribution to the energy density as Eq.~\eqref{eq:B0} plus higher derivative terms, and so it suffices to show that the ultraviolet contributions are subdominant to that from $B_0$.

To show this, we will first simplify the first factor in Eq.~\eqref{eq:I2BetaBar} by using Eqs.~\eqref{eq:kcLowerBound} and \eqref{eq:ScaleFactorPrime} and $\omega>k>k_c$ to show that $k^2m^2a^{\prime 2}\ll 16\omega^6$ and therefore $1-\frac{k^2m^2a^{\prime 2}}{16\omega^6}\approx 1$, provided the condition $a^2\ll B_0^{1/2}m^{-1}$ is satisfied, so we will need
\begin{equation}\label{eq:SpinorUltravioletConstraint1}
  a\ll B_0^{1/4}m^{-1} \;.
\end{equation}
This is the first of several conditions we will need.
We will consider the complete set of conditions collectively later.

With this simplification, one finds a bound on $I_2$ of
\begin{align}
  \abs{I_2} & \lesssim \int_{k_c}^{\infty}dk\left[ 2k^2\omega\magsq{\bar{\beta}_k} + \frac{1}{2}kma'\abs{\bar{\beta}_k} \right] + \frac{m^4a^{\prime 4}}{512k_c^4} \nonumber                                     \\
            & < \int_{k_c}^{\infty}dk\left[ 2k^3\magsq{\bar{\beta}_k} + km^2a^2\magsq{\bar{\beta}_k} + \frac{1}{2}kma'\abs{\bar{\beta}_k} \right] + \frac{m^4a^{\prime 4}}{512k_c^4}\;,  \label{eq:I2BoundStart}
\end{align}
where we have bounded the integral on the final term using the normalization~\eqref{eq:AlphaBetaNormalization} to obtain $\abs{\alpha_k} < 1$.
For $B_0$ to dominate, the contributions from $I_2$ must remain less than $B_0$, and therefore the full contribution from $I_2(\eta)=I_2(\eta_0)+\Delta I_2$ must be less than $B_0$.
We will find the conditions under which $\abs{I_2(\eta_0)}$ and $\abs{\Delta I_2}$, and therefore $\abs{I_2(\eta)}$, are subdominant to $B_0$.
%, which will imply that $\abs{I_2(\eta)}$ is indeed less than $B_0$ with the bound
%\begin{align}
%	\abs{I_2(\eta)} \leq \abs{I_2(\eta_0)} + \abs{\Delta I_2} \;.\label{eq:I2TotalContributionBound}
%\end{align}

Given Eq.~\eqref{eq:BetaBarBound}, one finds from Eq.~\eqref{eq:I2BoundStart} that the contributions from the integrand of $\abs{I_2(\eta_0)}$ are subdominant to $B_0$ provided that
\begin{subequations}
  \begin{align}
    A_0^2       & \ll B_0 \;,\label{eq:BetaBarSubdominantCondition1}       \\
    m^2a^2A_0^2 & \ll B_0^{3/2} \;,\label{eq:BetaBarSubdominantCondition2} \\
    ma'A_0      & \ll B_0 \label{eq:BetaBarSubdominantCondition3} \;.
  \end{align}
\end{subequations}
To satisfy~\eqref{eq:BetaBarBound}, one may increase $k_c$ which allows for decreasing $A_0$ and ensuring~\eqref{eq:BetaBarSubdominantCondition1} can be satisfied.
%Given Eq.~\eqref{eq:BetaBarBound}, one may decrease $A_0$ as $k_c$ increases, which ensures the first condition~\eqref{eq:BetaBarSubdominantCondition1} can be satisfied by choosing a large enough cutoff $k_c$.
Furthermore, provided Eq.~\eqref{eq:BetaBarSubdominantCondition1} is satisfied and using Eq.~\eqref{eq:ScaleFactorPrime}, one can show Eqs.~\eqref{eq:BetaBarSubdominantCondition2} and \eqref{eq:BetaBarSubdominantCondition3} become
\begin{align}
  a & \ll B_0^{1/4}m^{-1} \;,\label{eq:BetaBarSubdominantCondition4}   \\
  a & \ll B_0^{1/4}m^{-1/2} \;.\label{eq:BetaBarSubdominantCondition5}
\end{align}
The final term in~\eqref{eq:I2BoundStart} is also less than $B_0$ provided~\eqref{eq:BetaBarSubdominantCondition5} is satisfied.

For the bound on the integral in $\Delta I_2$ to converge, $\Delta\bar{\beta}_k$ must fall faster than $k^{-2}$.
We will assume that $\Delta\bar{\beta}_k$ falls at least as fast as $k^{-2-b_\Delta}$, and then we will need to show
\begin{equation}
  \abs{\Delta\bar{\beta}_k} < \frac{A_\Delta}{k^2}\left(\frac{k_c}{k}\right)^{b_\Delta} \;,\label{eq:DeltaBetaBarBound}
\end{equation}
with $A_\Delta$ independent of $k$ and $b_\Delta>0$ chosen appropriately for each term contributing to $\Delta\bar{\beta}_k$.
One finds from Eq.~\eqref{eq:I2BoundStart} that the contributions from $\Delta I_2$ are less than $B_0$ provided that
\begin{align}
  A_\Delta^2       & \ll B_0 \;,\label{eq:DeltaBetaBarSubdominantCondition1}       \\
  m^2a^2A_\Delta^2 & \ll B_0^{3/2} \;,\label{eq:DeltaBetaBarSubdominantCondition2} \\
  ma'A_\Delta      & \ll B_0 \label{eq:DeltaBetaBarSubdominantCondition3} \;.
\end{align}
One can obtain bounds on the contributions to $\Delta\bar{\beta}_k$, and hence the conditions under which Eqs.~\eqref{eq:DeltaBetaBarSubdominantCondition1}--\eqref{eq:DeltaBetaBarSubdominantCondition3} are satisfied, using the differential equation for $\bar{\beta}_k$.
From Eqs.~\eqref{eq:BetaDiffEq0thOrder} and \eqref{eq:BetaBar}, the differential equation is
\begin{align}
  \bar{\beta}_k'
   & = -\frac{ikma''}{4\omega^3}\alpha_k e^{-2i\theta_k} + \frac{ik^2m^2a^{\prime 2}}{8\omega^5}\beta_k + \frac{3ikm^3aa^{\prime 2}}{4\omega^5}\alpha_k e^{-2i\theta_k} \;. \label{eq:SpinorBetaBarDiffEq}
\end{align}
Writing $\bar{\beta}_k(\eta) = \bar{\beta}_k(\eta_0) + \Delta\bar{\beta}_k$, one then finds by integrating by parts on the phase
\begin{align}
  \Delta\bar{\beta}_k
  = & \int_{\eta_0}^{\eta}dx \left[ \frac{-ikma''}{4\omega^3}\alpha_ke^{-2i\theta_k} + \frac{ik^2m^2a^{\prime 2}}{8\omega^5}\beta_k + \frac{3ikm^3aa^{\prime 2}}{4\omega^5}\alpha_ke^{-2i\theta_k} \right] \nonumber                                                         \\
  = & \left.\frac{kma''}{8\omega^4}\alpha_ke^{-2i\theta_k}\right|_{\eta_0}^{\eta} + \int_{\eta_0}^{\eta}dx\,\bigg[ \frac{-kma'''}{8\omega^4}\alpha_ke^{-2i\theta_k} + \frac{3km^3aa'a''}{8\omega^6}\alpha_ke^{-2i\theta_k} + \frac{k^2m^2a'a''}{16\omega^6}\beta_k \nonumber \\
    & + \frac{ik^2m^2a^{\prime 2}}{8\omega^5}\beta_k + \frac{3ikm^3aa^{\prime 2}}{4\omega^5}\alpha_ke^{-2i\theta_k} \bigg] \;,
\end{align}
and therefore
\begin{align}
  \abs{\Delta\bar{\beta}_k}
  < & \frac{m\left(\abs{a''(\eta_0)} + \abs{a''(\eta)}\right)}{8k^3}
  + \frac{1}{8k^3} \int_{\eta_0}^{\eta}dx\left(m\abs{a'''} + m^2a^{\prime 2}\right) + \frac{1}{16k^4}\int_{\eta_0}^{\eta} dx\,m^2\abs{a'}\abs{a''} \nonumber     \\
    & + \frac{3}{4k^4}\int_{\eta_0}^{\eta}dx\,m^3a a^{\prime 2}+ \frac{3}{8k^5}\int_{\eta_0}^{\eta}dx\,m^3a\abs{a'}\abs{a''}  \;.\label{eq:DeltaBetaBarAbsValue}
\end{align}
Each term in Eq.~\eqref{eq:DeltaBetaBarAbsValue} can be written in the form of Eq.~\eqref{eq:DeltaBetaBarBound} to obtain conditions under which each will satisfy Eq.~\eqref{eq:DeltaBetaBarSubdominantCondition1}.
Term by term, using the inequalities in Eqs.~\eqref{eq:ScaleFactorPrime} and \eqref{eq:ScaleFactorPrime2}, we can obtain conditions on the scale factor under which each term is subdominant to the infrared contribution.
In order to obtain these conditions, we will assume $a'$, $a''$, and $a'''$ have definite signs; that is, each of them is either always positive or always negative.
The conditions are
\begin{subequations}
  \begin{align}
    \frac{ma''}{k^3} = \frac{ma''}{k^2k_c}\left(\frac{k_c}{k}\right)
    \implies & B_0\gg \left(\frac{ma''}{k_c}\right)^2 \nonumber                       \\
             & a\ll B_0^{1/4}m^{-1/3}\;, \label{eq:SpinorDeltaBetaBarBoundCondition1}
  \end{align}
  \begin{align}
    \frac{1}{k^3}\int dx\,m^2a'^2 < \frac{1}{k^3}\int dx\,m^2a^2a'
    = \frac{m^2a^3}{k^3}
             & = \frac{m^2a^3}{k^2k_c}\left(\frac{k_c}{k}\right) \nonumber \\
    \implies & B_0\gg \left(\frac{m^2a^3}{k_c}\right) \nonumber            \\
             & a\ll B_0^{1/4}m^{-2/3}\;,
  \end{align}
  \begin{align}
    \frac{1}{k^4}\int dx\,m^2a'a'' = \frac{m^2a'^2}{k^4}
             & = \frac{m^2a'^2}{k^2k_c^2}\left(\frac{k_c}{k}\right)^2 \nonumber \\
    \implies & B_0\gg \left(\frac{m^2a'^2}{k_c^2}\right)^2 \nonumber            \\
             & a\ll B_0^{1/4}m^{-1/2}\;,
  \end{align}
  \begin{align}
    \frac{1}{k^4}\int dx\,m^3aa'^2 < \frac{1}{k^4}\int dx\,m^3a^3a'
    = \frac{m^3a^4}{k^4}
             & = \frac{m^3a^4}{k^2k_c^2}\left(\frac{k_c}{k}\right)^2 \nonumber \\
    \implies & B_0\gg \left(\frac{m^3a^4}{k_c^3}\right)^2 \nonumber            \\
             & a\ll B_0^{1/4}m^{-3/4}\;,
  \end{align}
  \begin{align}
    \frac{1}{k^5}\int dx\,m^3aa'a'' < \frac{1}{k^5}\int dx\,m^3 a^4a'
    = \frac{m^3a^5}{k^5}
             & = \frac{m^3a^5}{k^2k_c^3}\left(\frac{k_c}{k}\right)^3 \nonumber        \\
    \implies & B_0\gg \left(\frac{m^3a^5}{k_c^3}\right) \nonumber                     \\
             & a\ll B_0^{1/4}m^{-3/5}\;. \label{eq:SpinorDeltaBetaBarBoundCondition5}
  \end{align}
\end{subequations}
Even if one of the three quantities $a'$, $a''$, or $a'''$ is not of definite sign, one can subdivide the integrals appearing in~\eqref{eq:DeltaBetaBarAbsValue} into regions in which it \textit{is} of definite sign, and assuming the number of regions is not too large the sum of these regions can be similarly bounded if the inequalities~\eqref{eq:SpinorDeltaBetaBarBoundCondition1}--\eqref{eq:SpinorDeltaBetaBarBoundCondition5} are satisfied.

We are working at times beyond the Planck era during which the mass is small compared to the Planck mass, and hence $m\ll 1$,
so the strongest restriction among Eq.~\eqref{eq:SpinorUltravioletConstraint1} and Eqs.~\eqref{eq:SpinorDeltaBetaBarBoundCondition1}--\eqref{eq:SpinorDeltaBetaBarBoundCondition5} for which the ultraviolet contributions are subdominant to that from $B_0$ is
\begin{equation}\label{eq:UltravioletSubdominantCondition}
  a \ll B_0^{1/4}m^{-1/3} \;.
\end{equation}

Equation~\eqref{eq:UltravioletSubdominantCondition} is stronger than the conditions in Eqs.~\eqref{eq:BetaBarSubdominantCondition4}--\eqref{eq:BetaBarSubdominantCondition5} and those in Eqs.~\eqref{eq:DeltaBetaBarSubdominantCondition2}--\eqref{eq:DeltaBetaBarSubdominantCondition3}.
One therefore has a complete set of conditions under which the ultraviolet contribution is subdominant to that from $B_0$: $k_c$ must be large enough such that Eq.~\eqref{eq:BetaBarSubdominantCondition1} is true, and $a$ must be small enough to satisfy Eq.~\eqref{eq:UltravioletSubdominantCondition}.

\subsubsection{Infrared Region\label{sec:InfraredRegion}}
It remains to be shown that $B_0$ is indeed the dominant contribution to the energy density among all the terms in the infrared contribution~\eqref{eq:I1}.
Given Eq.~\eqref{eq:B0}, the infrared contribution~\eqref{eq:I1} can be written
\begin{align}
  I_1 & = B_0 + \left(B-B_0\right) + \int_0^{k_c} dk\,\left[2k^2\left(\omega-k\right)\magsq{\beta_k} - \frac{k^4m^2a^{\prime 2}}{4\omega^5}\right] \;,\label{eq:I1b}
\end{align}
where we have defined
\begin{equation}
  B\equiv\int_0^{k_c}dk\,2k^3\magsq{\beta_k} \;.
\end{equation}

The $\left(B-B_0\right)$ term can be written
\begin{align}
  B-B_0 & = \int_0^{k_c}dk\,2k^3\left(\magsq{\beta_k} - \magsq{\beta_k(\eta_0)}\right) \nonumber                       \\
        & \leq \int_0^{k_c}dk\,2k^3\left(\magsq{\Delta\beta_k} + 2\abs{\beta_k(\eta_0)}\abs{\Delta\beta_k} \right) \;,
\end{align}
where $\Delta\beta_k = \beta_k - \beta_k(\eta_0)$.
From Eq.~\eqref{eq:BetaDiffEq0thOrder},
\begin{align}
  \abs{\Delta\beta_k} < \int_{\eta_0}^{\eta}dx\,\frac{kma'}{2\omega^2}
  =\frac{1}{2}\int_{\eta_0}^{\eta}dx\,\partial_x\tan^{-1}\left(\frac{ma}{k}\right)
  <\frac{ma}{2k} \;,
\end{align}
which implies
\begin{align}
  \int_{0}^{k_c}dk\,2k^3\magsq{\Delta\beta_k}
  < \frac{1}{4}k_c^2m^2a^2 \;\label{eq:SpinorI1DeltaB}
\end{align}
and
\begin{equation}
  \int_0^{k_c}dk\;4k^3\abs{\beta_k(\eta_0)}\abs{\Delta\beta_k}
  < \frac{2}{3}mak_c^3 \;,
\end{equation}
so, using Eq.~\eqref{eq:kcLowerBound}, one finds $\left(B-B_0\right)$ is dominated by $B_0$ if
\begin{equation}\label{eq:InfraredSubdominantCondition}
  a\ll B_0^{1/4}m^{-1} \;.
\end{equation}
Similarly, using the normalization condition~\eqref{eq:AlphaBetaNormalization} to bound $\magsq{\beta_k}\le1$, the next term in Eq.~\eqref{eq:I1b} is
\begin{align}
  \int_0^{k_c}dk\,2k^2\left(\omega-k\right)\magsq{\beta_k}
  \le \frac{1}{2}k_c^2m^2a^2 \;,\label{eq:SpinorI1BComplement}
\end{align}
which is also less than $B_0$ if $a\ll B_0^{1/4}m^{-1}$.
This condition is weaker than Eq.~\eqref{eq:UltravioletSubdominantCondition} and hence will be satisfied if the set of conditions under which the ultraviolet contribution is subdominant to $B_0$ is satisfied.

The last term in $I_1$ is
\begin{align}
  \int_0^{k_c}dk\,\frac{-k^4m^2a^{\prime 2}}{4\omega^5}
   & = \frac{m^2a^{\prime 2}}{4}\left(\frac{k_c^3}{3\omega_c^3} + \frac{k_c}{\omega_c}\right) + \frac{1}{4}m^2a^{\prime 2}\ln\left(\frac{ma}{\omega_c+k_c}\right) \nonumber \\
   & \approxeq \frac{1}{3}m^2a^{\prime 2} + \frac{1}{4}m^2a^{\prime 2}\ln\left(\frac{ma}{2k_c}\right) \;,\label{eq:SpinorI1Counterterm}
\end{align}
where we have used $a\ll B_0^{1/4}m^{-1}<k_c$ to simplify $\omega_c^2\equiv k_c^2+m^2a^2\approx k_c^2$.

\subsection{Friedmann Equation\label{sec:SpinorFieldsFriedmann}}
The energy density is given by the sum of the contribution from the infrared and ultraviolet regions.
As shown in Secs.~\ref{sec:UltravioletRegion}--\ref{sec:InfraredRegion}, provided one is looking early enough such that Eq.~\eqref{eq:InfraredSubdominantCondition} and hence Eq.~\eqref{eq:UltravioletSubdominantCondition} are satisfied, the energy density is dominated by the contributions from $B_0$ and the logarithm in Eq.~\eqref{eq:SpinorI1Counterterm}:
\begin{align}\label{eq:SpinorDominantEnergyDensityContribution}
  \rho_r \approxeq \frac{1}{\pi^2a^4}\left[ B_0 + \frac{1}{4}m^2a^{\prime 2}\ln\left(\frac{ma}{2k_c}\right) \right] \;.
\end{align}

Using the Friedmann equation
\begin{equation}\label{eq:SpinorFriedmann}
  \left(\frac{a'}{a^2}\right)^2 = \frac{8\pi}{3}\rho_r
\end{equation}
with the assumption that the contribution from massless spin-half fields dominates the energy density, one finds
\begin{equation}\label{eq:SpinorScaleFactorPrime}
  a^{\prime 2} \approxeq \frac{8B_0}{3\pi}\left[1-\frac{2}{3\pi}m^2\ln\left(\frac{ma}{2k_c}\right)\right]^{-1} \;.
\end{equation}
The renormalized energy density can be written using~\eqref{eq:SpinorScaleFactorPrime} as
\begin{equation}\label{eq:RenormalizedEnergyDensityFinal}
  \rho_r\approxeq \frac{B_0}{\pi^2a^4}\left\{1+\frac{2}{3\pi}m^2\left[\ln\left(\frac{2k_c}{ma_0}\right) - \ln\left(\frac{a}{a_0}\right)\right]\right\}^{-1} \;.
\end{equation}
At this point, we will choose $a_0\approx B_0^{1/4}$ to ensure we are past the Planck era, and given the condition~\eqref{eq:UltravioletSubdominantCondition} we have $a_1\approx B_0^{1/4}m^{-1/3}$, so $m^2\ln(a/a_0)<m^2\ln(m^{-1/3})\ll1$ is irrelevant compared to the $1$ term.
Hence, in the range $B_0^{1/4}<a<B_0^{1/4}m^{-1/3}$, the energy density will indeed be radiation dominated.

Note that $a'$~\eqref{eq:SpinorScaleFactorPrime} is approximately constant in the range of interest, so $a''$ will be strongly suppressed.
This allows one to relax some of the accumulated constraints, for example Eq.~\eqref{eq:SpinorDeltaBetaBarBoundCondition1}, implying $\rho\propto a^4$ over a larger range of $a$.
However, limits such as~\eqref{eq:InfraredSubdominantCondition} will generally not relax, but this is not surprising because at $a=B_0^{1/4}m^{-1}$ one expects fermions to become non-relativistic.
Of course, realistic models would not only have fermion fields but also an inflaton field, which would look like a cosmological constant, but, until the mass term becomes important at $a\sim B_0^{1/4}m^{-1}$ or inflation takes over, things will still be radiation dominated.

\section{Discussion and Conclusion\label{sec:DiscussionAndConclusion}}
In this paper we have analyzed the early-Universe, preinflationary behavior of massless vector fields of spin-1 and massive or massless fermion fields of spin-$\frac12$ in the semiclassical approximation.
We showed for a range of conformal time after the Planck era that both types of fields have radiation-dominated behavior.
Along with the same result obtained for scalar fields in~\cite{anderson_semiclassical_2020}, we have demonstrated that all matter fields which one might anticipate to play a role in the preinflationary era do in fact produce a radiation-dominated energy density that is typically assumed in inflationary models.

In Sec.~\ref{sec:VectorFields}, we summarized the adiabatic renormalization procedure for a massless vector field in a spatially flat FLRW universe, following the groundwork laid out in~\cite{chu_adiabatic_2017}.
We then used this procedure to renormalize the energy density contribution for such a field following a parameterization of the mode functions in terms of adiabatic states.
We found the renormalized energy density to have a radiation-dominated form similar to that for the scalar field in~\cite{anderson_semiclassical_2020}.

In Sec.~\ref{sec:SpinorFields}, we summarized a modified version of the adiabatic renormalization procedure for a massive or massless fermion field, using the modified WKB ansatz given in~\cite{landete_adiabatic_2013}.
We found that parameterizing the mode functions in terms of adiabatic states required higher than zeroth order contributions in order to properly renormalize the energy density at high energies.
We then made use of the leading second adiabatic-order contributions to the parameterization coefficients to obtain the leading order behavior of such high-energy contributions and used this to show the high-energy contributions are in fact subdominant to the radiation-like term in the remaining energy density contributions given a set of constraints on the scale factor.
These constraints are more stringent than those required for the scalar field~\cite{anderson_semiclassical_2020}, for which only a constraint on $(a^2)''$ was necessary.
We then found the next-to-leading order behavior of the energy density to be a logarithmic contribution and showed that it is subdominant compared to the radiation-dominated energy density for a the range of conformal time beyond the Planck era, where the constraint on the scale factor coming from the analysis of the high-energy contribution served as the upper bound on the range.

Our anaylsis of the fermion field assumed that fermionic matter was the dominant contribution to the energy density in order to make use of the Friedmann equation to obtain the approximate leading order behavior.
This result for fermions is different than that for scalars~\cite{anderson_semiclassical_2020} and vectors, which work in any case.
Our argument for fermions also only applies for the range $a_0<a<a_1$ described in Sec.~\ref{sec:SpinorFields}, though we expect this range to be in the preinflationary era prior to the fields becoming matter dominated.
This result is weaker than those for the other two fields, but in application it is not.
One would not be able to observe these matter fields during this era directly but rather through their effects on other phenomena, such as in the Cosmic Microwave Background, so the results are not necessarily weaker in application.

Having demonstrated in the semiclassical approximation that the matter content in the post-Planck, preinflationary era will indeed be radiation-dominated, one can proceed with the procedure for the inflaton field in~\cite{anderson_semiclassical_2020}.
There, it was assumed that spin-$\frac{1}{2}$ and spin-$1$ massive fields could be modeled with conformally coupled massive scalar fields, which were argued to be radiation-dominated themselves in the preinflationary era.
Our results that spin-$\frac{1}{2}$ fermion fields and massless spin-$1$ vector fields are radiation-dominated, themselves, supports the procedure in~\cite{anderson_semiclassical_2020}, and the analysis for obtaining a renormalized energy density for a universe with a mixture of cosmological constant and classical radiation, and hence for obtaining the power spectrum of the Cosmic Microwave Background, is identical.

\begin{acknowledgments}
  We would like to thank Paul Anderson for helpful insights and discussion.
\end{acknowledgments}

\bibliography{PreinflationaryRadiationDominatedFields}

\end{document}